\documentclass[american,aps,superscriptaddress,notitlepage,reprint]{revtex4-1}
\usepackage[T1]{fontenc}
\usepackage[latin9]{inputenc}
\usepackage{geometry}
\usepackage{array}
\usepackage{float}
\usepackage{multirow}
\usepackage{amsmath}
\usepackage{amssymb}
\usepackage{graphicx}

\makeatletter

\providecommand{\tabularnewline}{\\}

\makeatother

\usepackage{babel}
\begin{document}

\title{Stochastic method for calculating the ground state reduced density
matrix of trapped Bose particles in one dimension}

\author{Omri Buchman}

\affiliation{Fritz Haber Center for Molecular Dynamics, Institute of Chemistry,
The Hebrew University of Jerusalem, Jerusalem 91904, Israel}

\author{Roi Baer}
\email{roi.baer@huji.ac.il}

\affiliation{Fritz Haber Center for Molecular Dynamics, Institute of Chemistry,
The Hebrew University of Jerusalem, Jerusalem 91904, Israel}
\begin{abstract}
The reduced density matrix (RDM) is a fundamental contraction of the
Bose-Einstein condensate wave function, encapsulating its one-body
properties. It serves as a major analysis tool with which the condensed
component of the density can be identified. Despite its cardinal importance,
calculating the ground-state RDM of trapped interacting bosons is
challenging and has been fully achieved only for specific models or
when the pairwise interaction is weak. In this paper we discuss a
new approach to compute the RDM based on a double-walker diffusion
Monte Carlo random walk coupled with a stochastic permanent calculation.
We here describe the new method and study some of its statistical
convergence and properties applying it to some model systems.
\end{abstract}
\maketitle

\section{Introduction}

Despite its importance for determining the structure and properties
of trapped boson systems, calculation of the ground state RDM proves
to be a daunting task. It has been calculated exactly (near-analytically)
only for hard-core particles in harmonic traps \cite{Girardeau2001,Papenbrock2003}
and numerically-exactly for weakly interacting systems.\cite{Sakmann2008}
The ground state RDM of 3D trapped particles based on diffusion Monte
Carlo (DMC) with a variational Monte Carlo guiding function has been
used for studying systems of hard core bosons in 3D at various densities
with interactions of intermediate strength.\cite{dubois2003natural,Astrakharchik2004a}
The RDM in these approaches is evaluated by an approximate expression,
involving the variational and mixed estimators of the RDM but relying
quite significantly on the quality of the guiding function. This makes
the method inappropriate for strong interactions, where the approach
also tends to suffer from instabilities in the population control
resulting from singularities in the local energy under the guiding
function.\cite{DuBois2002} 

In this paper, we present a new stochastic approach for the calculation
of the ground-state RDM for trapped strongly interacting bosons. The
formalism seems applicable to any number of dimensions but in this
paper we describe and study the implementation to 1D bosons, which
are challenging systems due to their strong correlation effects.\cite{Petrov2000,Cazalilla2011}
The method is based on a DMC random walk and employs a stochastic
method for estimating the permanents required to calculate the RDM.
In section \ref{sec:Method} we describe the basic formalism, definitions
and techniques, in section \ref{sec:Applications} we apply the method
to systems of bosons trapped in a harmonic well where interaction
strength is increased while keeping the the trap potential fixed and
then in double-well traps where interaction strength is increased
while keeping the \emph{density }of the system (nearly) fixed; summary
and conclusions are given in Section~\ref{sec:Summary-and-Discussion}.

\section{\label{sec:Method}Method}

\subsection{\label{subsec:Basic-notions}Basic notions }

For $D$ bosons of mass $m_{b}$ in a trap potential $v\left(q\right)$
($q$ is the Cartesian position coordinate of a particle) interacting
through a pairwise potential $u\left(q_{12}\right)$, the Hamiltonian
is written as a sum of kinetic and potential energies:

\begin{align}
\hat{H} & =\hat{T}+\hat{V,}\label{eq:The Hamiltonian}\\
\hat{T} & =-\frac{\hbar^{2}}{2m_{b}}\sum_{n=1}^{D}\nabla_{n}^{2}\label{eq:The Kinetic Energy}\\
\hat{V} & =\sum_{n=1}^{D}v\left(q_{n}\right)+\sum_{m<n}^{D}u\left(\left|q_{n}-q_{m}\right|\right)\label{eq:The Potentia lEnergy}
\end{align}
where $\hat{V}$ is a sum of one-body and two-body interactions. Although
the formalism we develop is not limited to any specific form of the
trap potential or two body interactions, we will use, for demonstration
purposes, the following even-symmetric trap which combining a harmonic
well with a gaussian shaped barrier in its center:

\begin{equation}
v\left(q\right)=\frac{1}{2}m_{b}\omega^{2}q^{2}+V_{b}e^{-\frac{q^{2}}{2\sigma_{b}^{2}}}.\label{eq:TrapPotential}
\end{equation}
Here $\omega$, $V_{b}$ and $\sigma_{b}$ are, respectively, the
harmonic frequency, barrier height and barrier width. The interaction
we consider is a pairwise Gaussian repulsion of the form, 
\begin{align}
u\left(q_{12}\right) & =\frac{c}{\sqrt{2\pi}\sigma_{r}}e^{-\frac{q_{12}^{2}}{2\sigma_{r}^{2}}},\label{eq:InteractionPotential}
\end{align}
where $c$ is the repulsion strength and $\sigma_{r}$ the interaction
range. When addressing the purely harmonic trap ($V_{b}=0$) we will
use two pure quantities for characterizing the trap:
\begin{align}
\alpha_{0} & =\frac{c}{E_{0}l},\label{eq:alpha0}\\
\alpha_{1} & =\frac{\sigma_{r}}{l},\label{eq:alpha1}
\end{align}
where $E_{0}=\hbar\omega$ and $l=\sqrt{\frac{\hbar}{m_{b}\omega}}$
are the energy and length scales of the single particle non-interacting
harmonic ground-state. 

The ground-state reduced density matrix (RDM) for $D$ bosons is defined
up to a constant factor as an expectation value of a nonlocal operator:
\begin{align}
\Gamma_{1}\left(q,\tilde{q}\right) & \propto\int d\boldsymbol{x}d\tilde{\boldsymbol{x}}\Psi\left(\boldsymbol{x}\right)\Psi\left(\tilde{\boldsymbol{x}}\right)\times\label{eq:RDM-bas-def}\\
 & \,\,\,\,\,\,\,\delta\left(x_{1}-q\right)\delta\left(\tilde{x}_{1}-\tilde{q}\right)\prod_{j=2}^{D}\delta\left(x_{j}-\tilde{x}_{j}\right)\nonumber 
\end{align}
where $\Psi\left(\boldsymbol{x}\right)\equiv\Psi\left(x_{1},\dots,x_{D}\right)$
is the ground-state, symmetric to particle exchange and normalization
$\int\Gamma_{1}\left(q,q\right)dq=D$ can be imposed a posteriori.
Singling out particle 1 in this definition is arbitrary as all particles
are identical. In fact, we can take advantage of the wave function
exchange symmetry and write the RDM in an equivalent but explicitly
fully symmetric way: 

\begin{align}
\Gamma_{1}\left(q,\tilde{q}\right) & \propto\int d\boldsymbol{x}d\tilde{\boldsymbol{x}}\Psi\left(\boldsymbol{x}\right)\Psi\left(\tilde{\boldsymbol{x}}\right)\times\label{eq:RDM-sym-def}\\
 & \,\,\sum_{j}w\left(\boldsymbol{y}\left(\boldsymbol{x}|j\right),\boldsymbol{y}\left(\tilde{\boldsymbol{x}}|j\right)\right)\delta\left(x_{j}-q\right)\delta\left(\tilde{x}_{j}-\tilde{q}\right),\nonumber 
\end{align}
where $\boldsymbol{y}\left(\boldsymbol{x}|j\right)\equiv\left(y_{1},\dots,y_{D-1}\right)=\left(\dots,x_{j-1},x_{j+1},\dots\right)$
is the vector of $D-1$ coordinates obtained from the vector $\boldsymbol{x}$
by removing the $j$th coordinate. The weight $w\left(\boldsymbol{y},\tilde{\boldsymbol{y}}\right)$
of each double configuration $\boldsymbol{y}$, $\tilde{\boldsymbol{y}}$
is the number of permutations $P$ of the $\tilde{y}$ coordinates
having the property that simultaneously for all $k$, the position
$\tilde{y}_{P_{k}}$ is located in an infinitesimal volume element
surrounding the position of $y_{k}$. Mathematically this is expressed
as the following sum of products of delta-functions: $w\left(\boldsymbol{y},\tilde{\boldsymbol{y}}\right)\equiv\sum_{P}\prod_{k}\delta\left(y_{k}-\tilde{y}_{P_{k}}\right)$. 

For a numerical implementations, we coarse-grain the delta functions.
First, we introduce a q-axis grid containing $2N_{G}$ bins, each
of width $h$, centered on the grid points $y_{g}=gh$, where $g=-N_{G},-N_{G},+1,\dots,N_{G}-1,N_{G}$
is an integer. The coarse-grained RDM is then a histogram on a $2N_{G}\times2N_{G}$
lattice derived from the exact RDM as an integral over the bins:
\begin{equation}
\Gamma_{1}^{g\tilde{g}}\equiv h^{-2}\iint\Gamma_{1}\left(q,\tilde{q}\right)\theta_{h}\left(q-y_{g}\right)\theta_{h}\left(\tilde{q}-y_{\tilde{g}}\right)dqd\tilde{q},
\end{equation}
where $\theta_{h}\left(\xi\right)$ equals $1$ if $\xi\in\left[-\frac{h}{2},\frac{h}{2}\right]$
and zero otherwise. 

\begin{figure}[t]
\includegraphics[width=1\columnwidth]{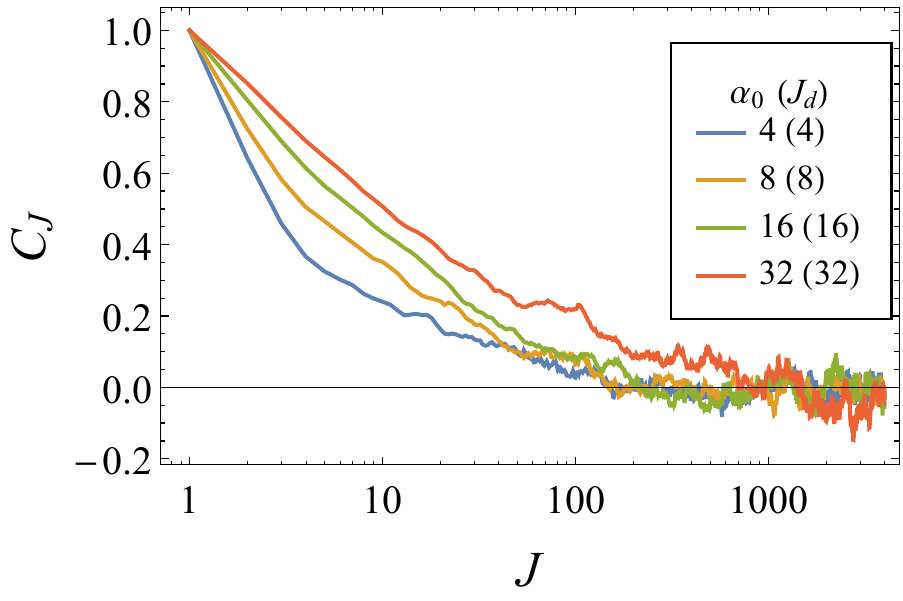}

\caption{\label{fig:ErefCorr}The autocorrelation function $C_{J}=\sum_{m}\left(E_{J+m}-\bar{E}\right)\left(E_{m}-\bar{E}\right)/\sum_{m}\left(E_{m}-\bar{E}\right)^{2}$,
where $E_{m}$ is the DMC reference energy at time-step $m$ and $\bar{E}$
is its mean, for $D=16$ bosons in a harmonic oscillator trap for
several values of $\alpha_{0}$ (and $\alpha_{1}=0.1$) . The decay
constant $J_{d}$ (defined by $C_{J_{d}}=e^{-1}\approx0.37$) is indicated
in parenthesis. Interestingly, in each case the value of $J_{d}$
is approximately equal to $\alpha_{0}$. }
\end{figure}

Next we introduce the DMC random walk as a means for calculating the
coarse grained RDM. Regular DMC produces a trajectory of length $N_{T}$
time steps made by $M$ walkers, giving $M\times N_{T}$ $D$-dimensional
vectors $\boldsymbol{x}$ distributed as the ground state wave function
$\Psi\left(\boldsymbol{x}\right)$. However, this is not what we need
for the RDM of Eq.~\ref{eq:RDM-sym-def}, where the integral is over
$\Psi\left(\boldsymbol{x}\right)\Psi\left(\tilde{\boldsymbol{x}}\right)$.
Hence we apply the standard DMC procedure not on a single but on a
double-walker system corresponding to $2\times D$ particles under
the Hamiltonian $\hat{H}=\hat{H}\left(\boldsymbol{x}\right)+\hat{H}\left(\tilde{\boldsymbol{x}}\right)$,
producing a random walk trajectory of $M\times N_{T}$ $2D$-dimensional
vectors $\left(\boldsymbol{x},\tilde{\boldsymbol{x}}\right)$ distributed
as the product of ground-state wave functions $\Psi\left(\boldsymbol{x}\right)\Psi\left(\tilde{\boldsymbol{x}}\right)$.
The coarse-grained RDM histogram then becomes equal (up to normalization)
to the following average along such a trajectory:
\begin{equation}
\Gamma_{1}^{g\tilde{g}}\propto\left\langle \sum_{j}w_{h}\left(\boldsymbol{y}\left(\boldsymbol{x}|j\right),\boldsymbol{y}\left(\tilde{\boldsymbol{x}}|j\right)\right)\theta_{h}\left(x_{j}-y_{g}\right)\theta_{h}\left(\tilde{x}_{j}-y_{\tilde{g}}\right)\right\rangle _{M\times N_{T}},\label{eq:RDM-calc}
\end{equation}
where, 
\begin{equation}
w_{h}\left(\boldsymbol{y},\tilde{\boldsymbol{y}}\right)=\sum_{P}\prod_{k}\theta_{h}\left(y_{k}-\tilde{y}_{P_{k}}\right)\label{eq:wh}
\end{equation}
are the coarse grained weights. The sum over the permutations is not
required when the random walk continues indefinitely, producing exhaustive
sampling (we can take $w_{h}\left(\boldsymbol{y},\tilde{\boldsymbol{y}}\right)=1$).
However sampling is evidently finite, and not taking the permutations
will result in extremely poor statistics because of the small probability
to find $y_{k}$ and $\tilde{y}_{k}$ in the same bin simultaneously
for all $k=1,\dots,D$. The sum of products over permutations appearing
in Eq.~\eqref{eq:wh} is the formal definition of a permanent of
the $\left(D-1\right)\times\left(D-1\right)$ matrix describing the
adjacency of particles in the two components of the double walker:

\begin{equation}
\Theta_{kj}=\theta_{h}\left(y_{k}-\tilde{y}_{j}\right).\label{eq:adjacencyMat}
\end{equation}
Note that the expression of the permanent in Eq.~\eqref{eq:wh} is
almost identical to that of the determinant except that in the latter
all odd permutations $P$ are multiplied by $-1$. Despite this similarity,
the numerical work needed to evaluate the permanent is vastly larger
than for the determinant: the former involves exponential complexity,
$O\left(2^{D}D\right)$ \cite{ryser1963combinatorial}, while the
latter is polynomial, $O\left(D^{3}\right)$. For this reason, we
use a stochastic method \cite{Godsil1981a} for evaluating the permanent
in polynomial time, as discussed in the following algorithm.

\subsection{\label{subsec:Algorithm-for-calculating}Algorithm for calculating
the reduced density matrix}

The $M$ DMC double walkers $\left(\boldsymbol{x}_{m},\tilde{\boldsymbol{x}}_{m}\right)$
($m=1,\dots,M$) are subject to the standard DMC diffusion and birth/death
processes in a series of $N_{T}$ time steps, each of duration $\Delta t$,
depending on the Hamiltonian $\hat{H}\left(\boldsymbol{x}\right)+\hat{H}\left(\tilde{\boldsymbol{x}}\right)$
as follows: 
\begin{enumerate}
\item \textbf{Diffusive step:} the ``position'' of each walker is changed
by $\left(\Delta\boldsymbol{x}_{m},\Delta\tilde{\boldsymbol{x}}_{m}\right)$,
a vector of random numbers, each sampled from the normal distribution
with mean $\mu=0$ and variance $\sigma^{2}=\frac{\hbar\Delta t}{m_{b}}$.
\item \textbf{Reproduction/annihilation:} For each walker at time $t$,
$\left(\boldsymbol{x}_{m},\tilde{\boldsymbol{x}}_{m}\right)$, an
integer $n=INT\left[e^{\left(E\left(t\right)-\left[V\left(\boldsymbol{x}_{m}\right)+V\left(\tilde{\boldsymbol{x}}_{m}\right)\right]\right)\Delta t/\hbar+M-M_{0}}+r\right]$
is computed (where $M_{0}$ is a preset target number of walkers),
$0\le r<1$ is a random fraction and $E\left(t\right)=\frac{1}{M}\sum_{k=1}^{M}\left[V\left(\boldsymbol{x}_{k}\right)+V\left(\tilde{\boldsymbol{x}}_{k}\right)\right]$
is the average potential energy over all walkers at time step $t=1,\dots,N_{T}$.
Then: 
\begin{enumerate}
\item if $n>0$ $n$ clones of the walker are generated and $M$ is increased
by $n$ 
\item if $n=0$ the walker is eliminated and $M$ is decreased by 1.
\end{enumerate}
\item \textbf{Evaluating the energy: }In the appropriate limit ($M\to\infty$,
$\Delta t\to0$ and $N_{T}\to\infty$) the expected time-step average
of $E\left(t\right)$ is an unbiased estimate of the ground state
energy of the double system: 
\begin{equation}
2E_{GS}=\left\langle \frac{1}{N_{T}}\sum_{n=1}^{N_{T}}E\left(n\Delta t\right)\right\rangle ,
\end{equation}
and the $M\times N_{T}$ walker positions $\left(\boldsymbol{x},\tilde{\boldsymbol{x}}\right)$
are distributed as $\Psi\left(\boldsymbol{x}\right)\Psi\left(\tilde{\boldsymbol{x}}\right)$.
The numerical procedure uses a finite number $M$ of walkers, a finite
time-step $\Delta t$ and a finite number of sampling times $N_{T}$,
leading estimates of $E_{GS}$ having random fluctuations $\Sigma_{\left(M,N_{T}\right)}\propto\frac{1}{\sqrt{N_{T}M}}$
as well as a small bias due to the finite time step $\Delta t$. 

\begin{figure}[t]
\includegraphics[width=1.1\columnwidth]{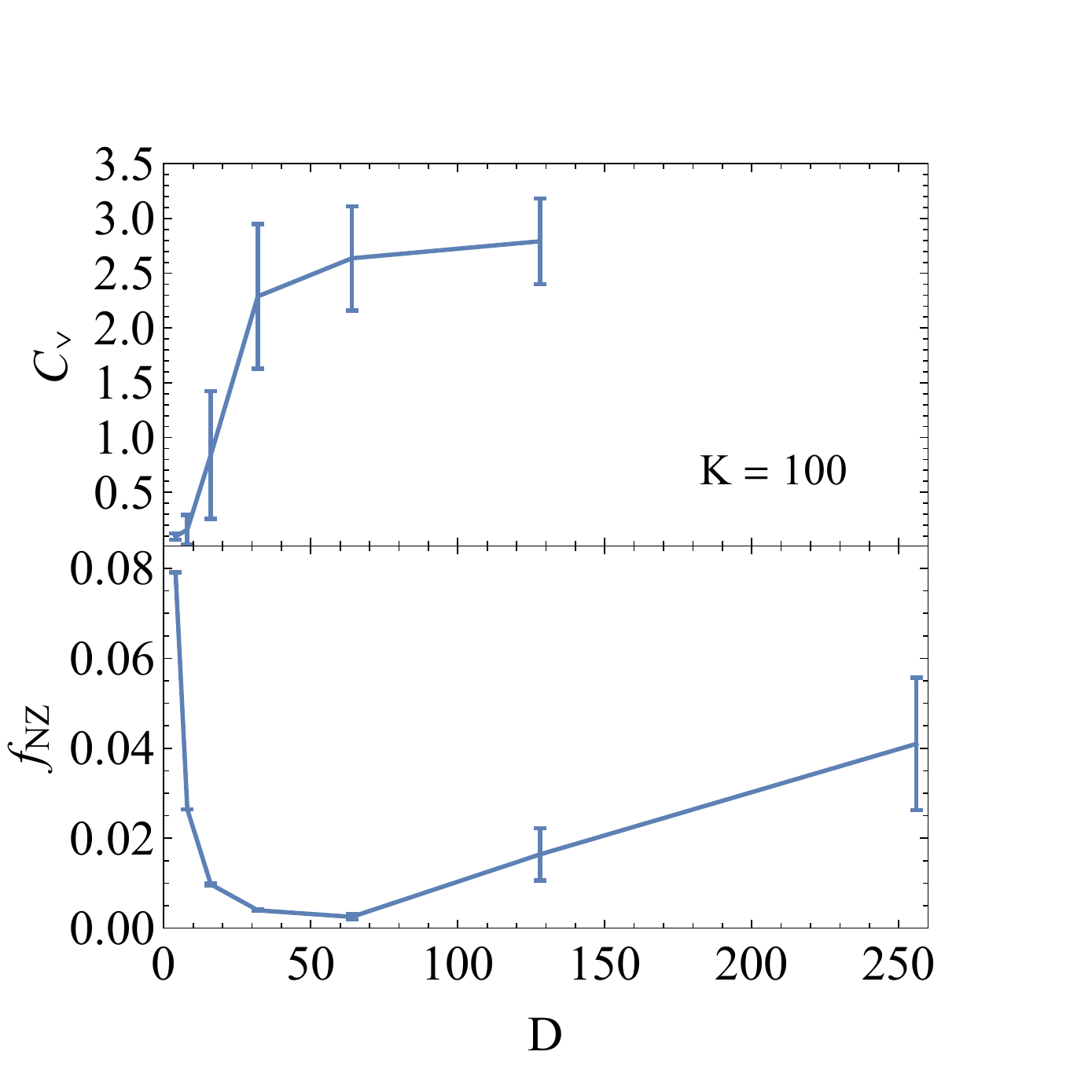}

\caption{\label{fig:GG-method}The application of the stochastic permanent
evaluation described in step \ref{enu:Stoch-perm} of the algorithm
in section \ref{subsec:Algorithm-for-calculating} to adjacency matrices
$\Theta_{ij}$ (Eq.~\ref{eq:adjacencyMat}) appearing in DMC trajectories
corresponding to $D$ interacting bosons inside a Harmonic well ($\alpha_{0}=4$,
$\alpha_{1}=0.1$ in Eqs.~\eqref{eq:alpha0}-\eqref{eq:alpha1}).
Top panel: The coefficient of deviation $C_{\nu}$ (relative standard
deviation) for the stochastic permanent evaluation as a function of
$D$. For each adjacency matrix $\Theta$, the permanent is reevaluated
stochastically 10 times (every time using $K=100$ sets of random
integers) and $C_{\nu}\left(\Theta\right)$ is calculated as the ratio
of the standard deviation to the average. The results shown in the
figure are averages $\left\langle C_{\nu}\left(\Theta\right)\right\rangle $
over 10000 instances of $\Theta$ matrices which arise during the
DMC random walk. Bottom panel: The frequency of non-zero permanents
as a function of $D$. }
\end{figure}

\item \textbf{Estimating the RDM:} Every $N_{C}$ time steps the DMC double
walkers are used update the RDM histogram according to Eq.~\eqref{eq:RDM-calc}.
$N_{C}$ is taken much larger than the the correlation decay lengths
$J_{d}$ of the walk (see Fig.~\ref{fig:ErefCorr}). In Eq.~\eqref{eq:RDM-calc},
the bosonic weight $w_{h}\left(\boldsymbol{y},\tilde{\boldsymbol{y}}\right)$
is equal to the permanent of the $\left(D-1\right)\times\left(D-1\right)$
adjacency matrix $\Theta_{ij}$ of Eq.~\eqref{eq:adjacencyMat} which
is evaluated following these steps:
\begin{enumerate}
\item \label{enu:Prelim-screen}Preliminary screening: we compute the column
sums $c_{j}=\sum_{i=1}^{D}\Theta_{ij}$ and the row sums $r_{i}=\sum_{j=1}^{D}\Theta_{ij}$
of the adjacency matrix and if one of these is zero the permanent
is immediately set to zero without further computation. The numerical
effort in this screening process scales at most as $O\left(D^{2}\right)$
and is effective since typically only a small fraction of the permanents
are nonzero (see bottom panel of Fig.~\ref{fig:GG-method}).
\item \label{enu:Stoch-perm}For the adjacency matrices passing step~\ref{enu:Prelim-screen},
the permanent is estimated as the average $\left\langle \left|\det\Phi\right|^{2}\right\rangle $
where $\Phi$ is the matrix obtained from $\Theta$ by multiplying
each of its elements by $\pm1$ at random. Mathematically, $\Phi_{ij}=\left(-\right)^{n_{ij}}\Theta_{ij}$
where $n_{ij}$ are random independent integers.\cite{Godsil1981a}
The average $\left\langle \left|\det\Phi\right|^{2}\right\rangle $
is estimated using $K$ samples of the integers $n_{ij}$, where $K$
is on the order of a few hundreds. The relative standard deviation
$C_{\nu}$ occurring in this stochastic permanent evaluation for a
typical DMC trajectory is shown in the top panel of Fig.~\ref{fig:GG-method}
for $K=100$. $C_{\nu}$ grows roughly in proportion to $D$, for
large $D$'s. 
\item Normalize ($\Gamma_{1}^{g\tilde{g}}\leftarrow\Gamma_{1}^{g\tilde{g}}\times\frac{D}{tr\Gamma_{1}h^{2}}$)
and symmetrize ($\Gamma_{1}^{g\tilde{g}}\leftarrow\left(\Gamma_{1}^{g\tilde{g}}+\Gamma_{1}^{\tilde{g}g}\right)/2$)
the completed RDM histogram of Eq.~\eqref{eq:RDM-calc}.
\end{enumerate}
\end{enumerate}
We found the statistical error $\Sigma_{RDM}$ of any RDM property
we looked at (eigenvalues, for example) is proportional to $\frac{1}{\sqrt{N_{T}MK}}$
where $N_{T}$ is the number of time steps, $M$ the number of walkers
and $K$ the number of determinants used in the permanent evaluation.
From this, we conclude that the bias, if it exists, is small and the
error is dominated by statistical fluctuations.

The algorithm quickly identifies most of the zero permanents however
it is clear that for the sampling to be efficient we cannot afford
a situation where the permanents are rarely different than zero. Hence,
the bin size $h$ should not be too small, and a general rule of the
thumb would be to take $h$ to be of the order of $n^{-1}$ (or a
small fraction thereof) where $n$ is the average density. The efficacy
of the permanent method is seen in that the fraction of non-zero permanents
grows with increasing number of particles for a Harmonic traps (see
bottom panel of Fig.~\ref{fig:GG-method}). This finding has support
of theoretical investigations.\cite{Tao2009} Thus, the sampling efficiency
is not expected to decrease and perhaps even increases as the number
of particles grows.

\begin{figure}[t]
\includegraphics[width=0.99\columnwidth]{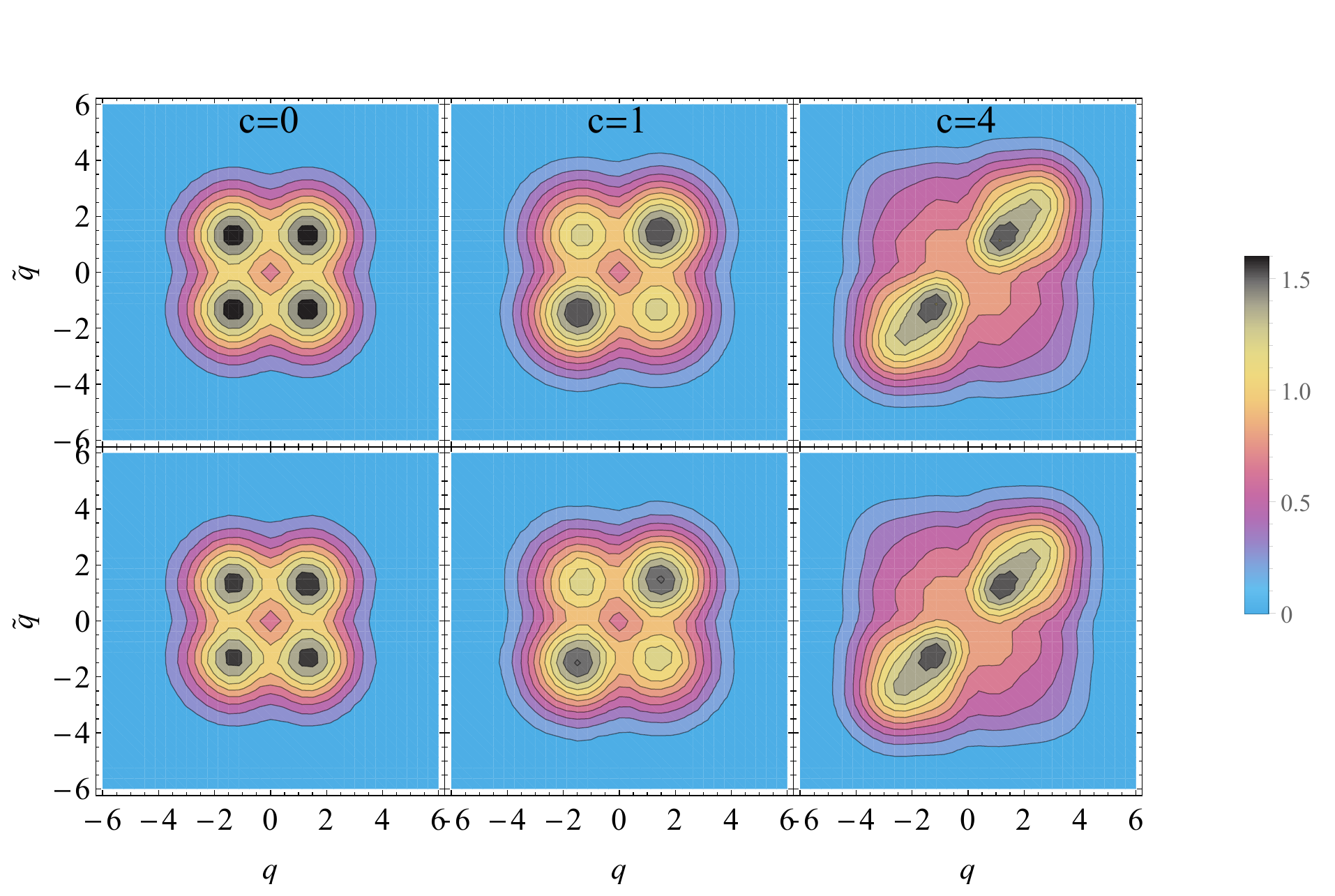}
\centering{}\caption{\label{fig:DM-Validation}The deterministic (top panels) and DMC (bottom
panels) RDM $\Gamma_{1}\left(q,\tilde{q}\right)$ for of $D=4$ unit-mass
particles trapped in the potential well $v\left(x\right)$ of Eq.~\ref{eq:TrapPotential}
and interacting via the potential $u\left(x_{12}\right)$ of Eq.~\ref{eq:InteractionPotential}.
The parameters are $k_{H}=0.25$, $V_{b}=1.5$, $\sigma_{b}=0.5$
and $\sigma_{r}=0.5$ and the values of $c$ are indicated in the
figure for each column. A bin-size of $h=0.375$ atomic units was
used for sampling. The DMC calculation used a total of $M=48000$
walkers ($128000$ for $c=4$), $N_{T}=4000$ time steps ($20000$
for $c=4$) and $\Delta t=0.01$ time units ($0.005$ for $c=4$). }
\end{figure}

\subsection{\label{subsec:Statistics-and-validation}Statistics and validation}

In Fig.~\ref{fig:DM-Validation} we show contour plots of a grid-based-deterministic
and the DMC-based-stochastic RDM estimates of $\Gamma_{1}\left(q,\tilde{q}\right)$
for several systems of $D=4$ particles interacting with increasing
repulsion strengths. In each case the DMC-based and grid-based RDMs
are indeed nearly identical in appearance, due to extensive sampling,
validating in principle, our method.

To show the effect of the stochastic permanent evaluation, we study
the three highest-lying RDM eigenvalues for a set of 16 bosons in
a Harmonic trap, as shown in Table~\ref{tab:RDM-Statistics}. The
averages and fluctuations using DMC with deterministic permanent evaluation
and DMC with stochastic permanent evaluation for $K=200$ and $400$
stochastic determinants are shown. The expectation values are close
and the standard deviations with $K=400$ are close to the deterministic
fluctuations. 

\begin{table}[H]
\begin{centering}
\begin{tabular}{|c|c|c|c|c|c|c|}
\hline 
$K$ & \multicolumn{2}{c|}{$0$} & \multicolumn{2}{c|}{$200$} & \multicolumn{2}{c|}{$400$}\tabularnewline
\hline 
 & $E\left(f\right)$ & $\sigma\left(f\right)$ & $E\left(f\right)$ & $\sigma\left(f\right)$ & $E\left(f\right)$ & $\sigma\left(f\right)$\tabularnewline
\hline 
$f_{1}$ & $0.806$ & $0.01$ & $0.808$ & $0.014$ & $0.805$ & $0.008$\tabularnewline
\hline 
$f_{2}$ & $0.085$ & $0.006$ & $0.086$ & $0.009$ & $0.085$ & $0.007$\tabularnewline
\hline 
$f_{3}$ & $0.047$ & $0.006$ & $0.043$ & $0.003$ & $0.047$ & $0.006$\tabularnewline
\hline 
\end{tabular}
\par\end{centering}
\caption{\label{tab:RDM-Statistics}The expected value and standard deviation
of the 3 largest RDM eigenvalues for a system of 16 bosons in a Harmonic
trap, calculated using DMC comparing the deterministic $(K=0$) and
stochastic ($K=200$, $400$) evaluations of permanents. The parameter
$K$ is the number of stochastic determinant calculations used for
each permanent evaluation. The potential parameters (see Eqs.~\eqref{eq:alpha0}-\eqref{eq:alpha1})
are $\alpha_{0}=4$, $\alpha_{1}=0.1$. The DMC calculation used $M=64000$
walkers and $N_{T}=8000$ time steps with $\Delta t=0.005\omega^{-1}$
and the RDM bin size was $\Delta x=0.625l$. }
\end{table}

\begin{figure*}[t]
\includegraphics[width=0.65\textheight]{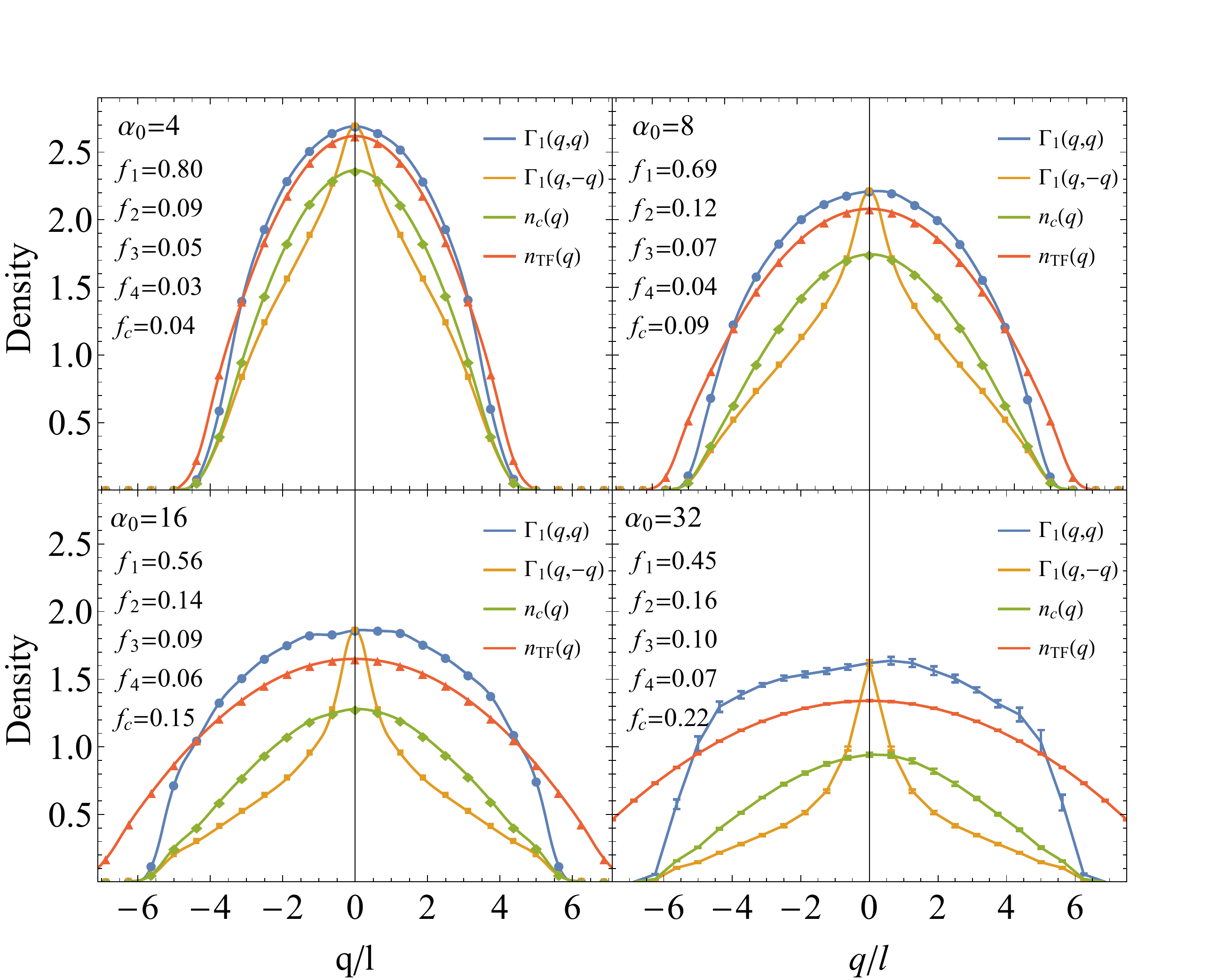}

\caption{\label{fig:hWellRDM}The RDM diagonal, anti-diagonal, condensate ($n_{C}\left(q\right)$)
and Thomas-Fermi ($n_{TF}\left(q\right)$) densities for $D=16$ particles
in a Harmonic well interacting via the potential of Eq.~\eqref{eq:InteractionPotential}.
The interaction range parameter is $\alpha_{1}=0.1$ while the interaction
strength parameter $\alpha_{0}$ is indicated in the panels. The RDM
eigenvalues (divided by $D$) are the eigenstate fractions $f_{n}$
indicated in each panel with $f_{1}$ and $f_{2}$ having relative
errors of $10\%$ and $f_{3}$ and $f_{4}$ of $20\%$ (largely independent
of $\alpha_{0}$). For $\alpha_{0}<32$ the statistical error bars
are not larger than the marker symbols. For $\alpha_{0}=32$ the statistical
error bars are shown explicitly for the diagonal, antidiagonal and
condensate densities. The RDM bin size was $h=0.625l$.}
\end{figure*}

\section{\label{sec:Applications}Applications}

In this section we apply the algorithm for two types of trapped boson
systems in order to demonstrate the performance and the kind of results
that can be obtained. We compare the calculated densities to that
of the Thomas-Fermi (TF) approximation \cite{Thomas1927,Fermi1927},
given as the \emph{positive part }of the shifted and negatively-scaled
potential well:
\begin{equation}
n_{TF}\left(q\right)=\mathcal{P}\frac{\mu-v\left(q\right)}{c}.\label{eq:TF-density}
\end{equation}
Here, the TF chemical potential $\mu$ is determined by the density
normalization condition $\int n_{TF}\left(q\right)dq=D$. 

\subsection{\label{subsec:Harmonic-Well}Constant harmonic-well trap}

\begin{table}[b]
\begin{tabular}{|l|c|c|c|c|}
\hline 
\multirow{2}{*}{DMC run data} & \multicolumn{4}{c|}{$\alpha_{0}$}\tabularnewline
\cline{2-5} 
 & $4$ & $8$ & $16$ & $32$\tabularnewline
\hline 
$M$ ($\times10^{3}$) & $96$ & $96$ & $480$ & $640$\tabularnewline
\hline 
$N_{T}$ ($\times10^{3}$) & $900$ & $900$ & $900$ & $6000$\tabularnewline
\hline 
$N_{J}$ & $250$ & $250$ & $500$ & $500$\tabularnewline
\hline 
$K$ & $100$ & $100$ & $100$ & $100$\tabularnewline
\hline 
$\omega\Delta t$ ($\times10^{-3}$) & $5$ & $2.5$ & $2.5$ & $2.5$\tabularnewline
\hline 
Wall time $hrs\times$CPU & $11\times3$ & $10\times3$ & $33\times6$ & $192\times8$\tabularnewline
\hline 
\end{tabular}

\caption{\label{tab:hWell-params}The parameters for the DMC runs used to produce
the results shown in Fig.\ref{fig:hWellRDM}. The wall time in hours
and the number of core-i7 CPU's used (each CPU running 8 threads).}
\end{table}

In Fig.~\ref{fig:hWellRDM} we study 16 trapped bosons in a harmonic
well as a function of $\alpha_{0}$, taking the values $4$, $8$,
$16$, $32$ with $\alpha_{1}=0.1$ and $V_{b}=0$ (corresponding
DMC run parameters given in Table~\ref{tab:hWell-params}). We choose
the regime of small $\alpha_{1}$ so the interaction is close to ``contact''.
A useful way to think of this series of systems is to imagine that
the repulsion strength $c$ increases (in proportion to $\alpha_{0}$)
while the harmonic trap stays put. 

\begin{figure*}[t]
\includegraphics[width=0.65\textheight]{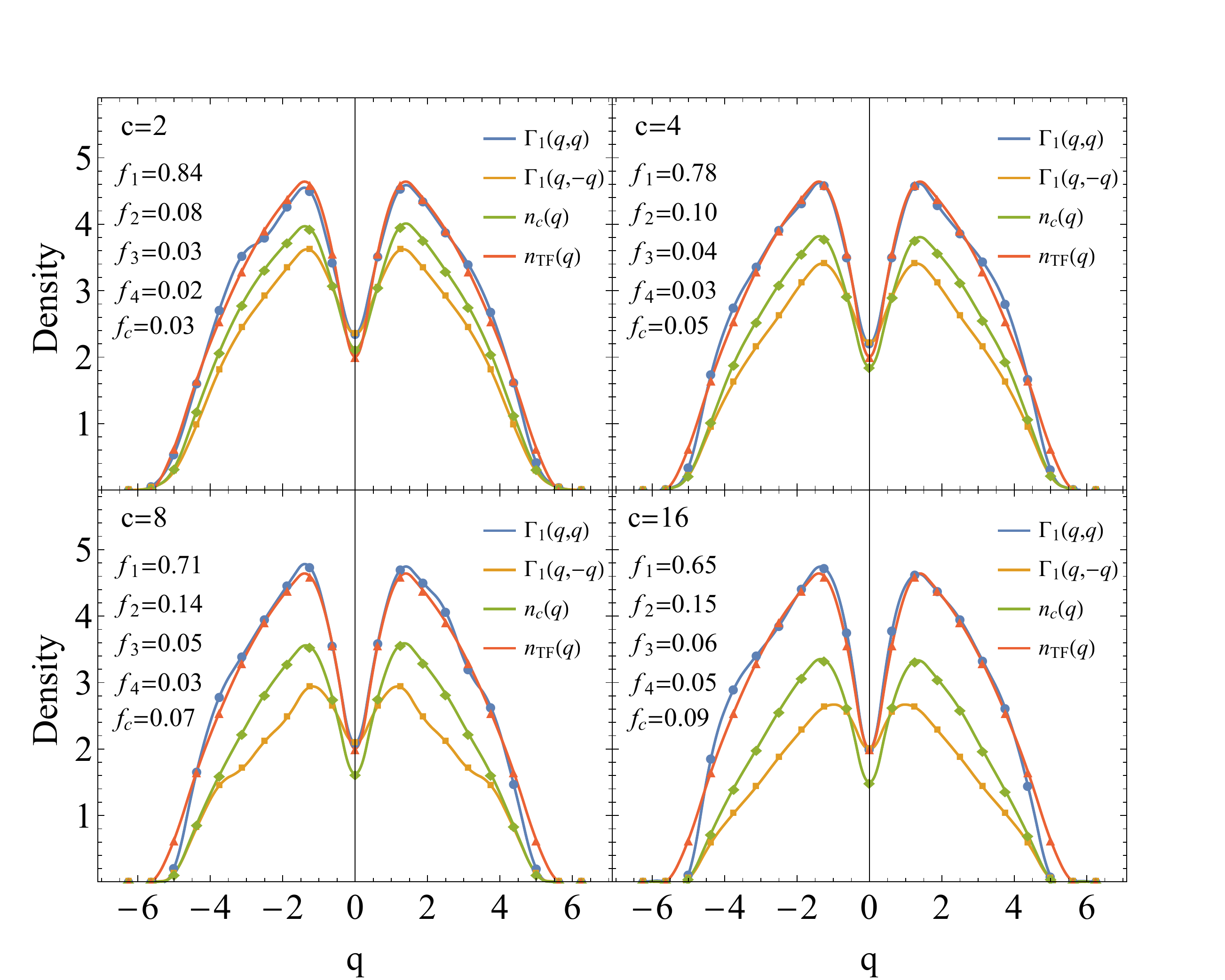}

\caption{\label{fig:dblWellRDM}The RDM diagonal, anti-diagonal, condensate
($n_{C}\left(q\right)$) and Thomas-Fermi ($n_{TF}\left(q\right)$)
densities for $D=32$ particles with the value of $c$ indicated in
the panels and $\sigma_{r}=0.1$, $\sigma_{b}=0.5$. The other two
potential parameters are taken as $k_{H}=2.86c$ and and $V_{b}=3c$.
This forces the TF density to be identical in all four systems. The
RDM eigenstate fractions $f_{n}$, $n=1,\dots,4$ as well as $f_{c}=\sum_{n>4}f_{n}$
, are indicated in each panel; $f_{1}$ and $f_{2}$ have relative
errors of $10\%$ and $f_{3}$ and $f_{4}$ of $20\%$ (largely independent
of $c$). The DMC parameters are given in Table~\ref{tab:dblW-params}.
The bin size was $h=0.625$ length units. }
\end{figure*}

It is seen that as the repulsion ($\alpha_{0}$) grows, the density
diminishes and broadens. This happens because at short inter particle
distances the repulsion force is stronger than the harmonic force
and thus, as repulsion grows the particles can stretch the harmonic
spring and spread out. 

For $\alpha_{0}=4$ and $8$ the density $\Gamma_{1}\left(q,q\right)$
in Fig.~\ref{fig:hWellRDM} is similar in shape to the TF density
$n_{TF}\left(q\right)$ (Eq.~\ref{eq:TF-density}). The TF approximation
is expected to apply for large numbers of particles,\cite{Petrov2000}
and weak interactions $\alpha_{0}\ll1$, and is seen here to work
surprisingly well beyond this limit. As $\alpha_{0}$ increases further,
the system gradually assumes a more Fermionic structure, which includes
a flattening of the density profile. But the TF density retains the
parabolic shape and therefore is not any more a reasonable approximation
to the density. 

As for the condensate density $n_{c}\left(q\right)$. For the lowest
value $\alpha_{0}=4$, it is very similar in shape to the total density,
just scaled by a factor $f_{1}\approx0.8$ where $f_{1}$ is the condensate
fraction. As $\alpha_{0}$ increases the condensate is gradually destroyed.
This is evidence by the steady decrease of the condensate fraction
$f_{1}$ and then by the anti-diagonal density $\Gamma_{1}\left(q,-q\right)$,
progressively developing a concave shorter ranged character while
deviating in shape from the total density $\Gamma_{1}\left(q,q\right)$
(see Appendix~\ref{sec:RDM-Diagonal-and-antidiagonal} for discussion).
Finally as interactions grow, the shape of the condensate density
$n_{c}\left(q\right)$, retaining its flexible smoothness, increasingly
deviates from that of the total density which displays increasing
rigidity due to fermionization. 

Fig.~\ref{fig:hWellRDM} also displays the statistical error bars
for the $\alpha_{0}=32$ system. It is seen that the total density
is considerably more sensitive to the QMC statistical fluctuations
than the condensate density (and the anti-diagonal density). This
is reminiscent of the two-fluid model of superfluid He-II \cite{Tisza1938}
according to which the condensate has vanishing viscosity and therefore
is immune to fluctuations quite distinct from the behavior of the
normal fluid.\footnote{The viscosity of a fluid is related momentum fluctuations by the Green-Kubo
formula.}

\begin{table}[b]
\begin{tabular}{|l|c|c|c|c|}
\hline 
\multirow{2}{*}{DMC run data} & \multicolumn{4}{c|}{$c$}\tabularnewline
\cline{2-5} 
 & $2$ & $4$ & $8$ & $16$\tabularnewline
\hline 
$M$ ($\times10^{3}$) & $256$ & $256$ & $256$ & $512$\tabularnewline
\hline 
$N_{T}$ ($\times10^{3}$) & $75$ & $140$ & $250$ & $350$\tabularnewline
\hline 
$N_{J}$ & $50$ & $100$ & $250$ & $500$\tabularnewline
\hline 
$K$ & $100$ & $100$ & $100$ & $100$\tabularnewline
\hline 
$N_{det}=MN_{T}K/N_{J}$ ($\times10^{10}$) & $3.8$ & $3.6$ & $2.6$ & $3.6$\tabularnewline
\hline 
$\omega\Delta t$ ($\times10^{-3}$) & $1.25$ & $1.25$ & $1.25$ & $1.25$\tabularnewline
\hline 
Wall time $hrs\times$CPU & $64\times4$ & $65\times4$ & $53\times4$ & $47\times8$\tabularnewline
\hline 
\end{tabular}

\caption{\label{tab:dblW-params}The parameters for the DMC runs ($D=32$ bosons
in a double well, keeping the density constant as the interaction
constant $c$ grows) used to produce the results shown in Fig.\ref{fig:dblWellRDM}.
The wall-time in hours and the number of core-i7 CPU's used (each
CPU running 8 threads). The DMC correlation time for $c=16$ was large
and required large $N_{J}$ to reduce fluctuations. }
\end{table}

\subsection{\label{subsec:Double-well-trap-with}Constant density in double-well
trap }

The generality of the DMC-based RDM calculation allows us to study
systems beyond the uniform gas and the harmonic trap approximations.
One interesting case, is the partially-fragmented trapped gas, which
is formed in a double-well potential. When the barrier is extremely
wide and tall, the system fragments into two condensates \cite{Noziere1995,Mueller2006}
with RDM exhibiting two large and equal eigenvalues. However, if the
barrier is only partially separating the condensate the nature of
the system is mixed and difficult to describe without detailed calculation. 

Here we examined the behavior of the bosons when trapped in a double
well as the repulsion strength is increased. If we keep the trap constant
and just increase the repulsion we find that the effect of the constant
barrier becomes negligible and the systems gradually shifts towards
that of bosons trapped in a harmonic well. In order to prevent this,
we examine systems of increasing repulsion constant $c$ and at the
same time but we change the trap (spring constant $k_{H}$ and barrier
height $V_{b}$ in Eq.~\eqref{eq:TrapPotential}) so that the boson
density stays (nearly) constant. This is a different limit than that
studied in the previous section, where we kept the trap constant as
we increased $c$ and the density decreased. We found that with constant
$\sigma_{r}=0.1$ and $\sigma_{b}=0.5$, the TF density is unchanged
if we preserve the ratios $V_{b}/c$ and $k_{H}/c$ (we took these
equal to 3 and 2.86 respectively). The RDM properties of 4 such systems
with $c=2$, $4$, $8$, and $16$ are shown in \eqref{fig:dblWellRDM},
(corresponding DMC run parameters given in Table~\ref{tab:dblW-params}).
Since the density is kept constant the main response is expressed
as off diagonal changes in the RDM as $c$ grows. What we see is that
the anti-diagonal $\Gamma_{1}\left(q,-q\right)$ gradually diminishes
for intermediate values of $q$ and deforms, smearing the double-hump
feature. The condensate density, like the total density $\Gamma_{1}\left(q,q\right)$,
seems to preserve it's shape but reduces as contributions from other
eigenfunctions of the RDM grow. Indeed, the strengthening of $c$
reduces the value of the condensate fraction, i.e. the largest RDM
eigenvalue fraction, from $f_{1}=0.84$ at $c=2$ to $f_{1}=0.65$,
while compensating by increasing the other eigenvalue fractions $f_{2}$,
$f_{3}$ and $f_{4}$. Note that the growing value of the sum of higher
state population fractions $f_{c}=\sum_{k>4}f_{k}$, reaching 9\%
at $c=16$. The second eigenvalue does not grow appreciably larger
than the third or fourth eigenvalue fractions, showing that the condensate
is not ``fragmented'' despite the visibly deep cut through the density
at $x=0$.

\section{\label{sec:Summary-and-Discussion}Summary and Discussion}

In this paper we have developed a new stochastic method for calculating
the RDM of trapped Bose particles in the ground state. The method
is based on a unguided DMC process in which a double-walker is used
to estimate the RDM $\Gamma_{1}^{q\tilde{q}}$ (where $q$ designate
bins on the position axis) as a permanent of the double-walker adjacency
matrix. We have used the method to treat systems of up to 32 bosons
with usefully converged statistics in harmonic and double-well traps.
Based on the tests we ran, we estimate the complexity to scale as
$D^{6}=D^{3}\times D^{2}\times D$ where the first factor is due to
the complexity of a determinant calculation, the second is our estimate
of the increase in the number of determinant evaluations needed for
each permanent calculation due to the linear increase of the relative
statistical fluctuations $C_{\nu}$ with $D$ (top panel of Fig.~\ref{fig:GG-method})
and the third is due to the fact that for each double walker we repeat
the permanent evaluation $D$ times. In a limited range of $D$, the
efficiency of the sampling decreases with increasing $D$ due to the
decrease in the number of non-zero permanents (see the bottom panel
of the figure). However, when $D$ grows further this effect will
diminish since the fraction of non-zero permanents actually grows
with $D$. In calculating the RDM of harmonically trapped particles
with $\alpha_{0}=4$ and $\alpha_{1}=0.1$, the CPU time increased
by a factor \textasciitilde{}50 (keeping the same level of statistical
fluctuations) when going from $D=16$ to $D=32$, which is consistent
with this scaling. Note however, that this estimated complexity is
based on experience with the Harmonic-trapped Bosons and short interaction
ranges. Its generality needs to be further investigated tested in
different settings and applications. 

We point out that while in this paper we focused on short ranged repulsive
1D particles, there is no formal reason why the method will not be
applicable for higher dimensions and other types of interactions.
Indeed the possibility of these issues is left as future directions.

It is important to appreciate, that the present stochastic RDM calculation
essentially involves a stochastic post-processing step placed \emph{on
top }of a DMC random walk. As such, the same technique can perhaps
be used in conjunction with other types of Monte Carlo methods or
even with deterministic approaches that produce a wave function. This
too is a possible direction for extending the method.

\appendix

\section{\label{sec:RDM-Diagonal-and-antidiagonal}RDM Diagonal and anti-diagonal
for potentials with inversion symmetry}

The condensate is associated with the antidiagonal long range of the
density matrix.\cite{landau1941theory,Penrose1956} In finite systems
it is more difficult to speak of long range yet the relation, e.g.
ratio, of the anti-diagonal and diagonal can be considered. We describe
this approach here.

For the RDM of the (non-negative) ground-state, as considered here,
the RDM $\Gamma_{1}\left(q,\tilde{q}\right)$ is also manifestly non-negative.
Furthermore, it the trap potential is symmetric $v\left(q\right)=v\left(-q\right)$,
the RDM eigenstates $\psi_{n}\left(q\right)$ ($\Gamma_{1}\left(q,\tilde{q}\right)=\sum_{n}w_{n}\psi_{n}\left(q\right)\psi_{n}\left(\tilde{q}\right)$
where $1\ge w_{n}\ge0$ are the RDM eigenvalues)are either symmetric
or antisymmetric to inversion. The diagonal and antidiagonal densities
can thus be written as
\begin{align}
\Gamma_{1}\left(q,q\right) & =\sum_{n}w_{n}\left|\psi_{n}\left(q\right)\right|^{2}\\
\Gamma_{1}\left(q,-q\right) & =\sum_{\psi\in even}w_{n}\left|\psi_{n}\left(q\right)\right|^{2}-\sum_{\psi\in odd}w_{n}\left|\psi_{n}\left(q\right)\right|^{2}
\end{align}
Focusing on the sum and difference between the RDM diagonal $\Gamma_{1}\left(q,q\right)$
and anti-diagonal $\Gamma_{1}\left(q,-q\right)$, we define two non-negative
even $\left(+\right)$ and odd $\left(-\right)$ state densities 
\begin{equation}
n_{\pm}\left(q\right)=\frac{1}{2}\left(\Gamma_{1}\left(q,q\right)\pm\Gamma_{1}\left(q,-q\right)\right),\label{eq:diag-antidiag}
\end{equation}
and the corresponding even/odd populations $D_{\pm}=\int n_{\pm}\left(q\right)dq$.
Clearly, the sum $D_{+}+D_{\text{-}}=\int\Gamma_{1}\left(q,q\right)dq$
is the total population $D$, while the difference,

\begin{equation}
D_{+}+D_{\text{-}}=\int\Gamma_{1}\left(q,-q\right)dq\label{eq:anti-diag-integral}
\end{equation}
is the integral of the anti-diagonal (which is thus always positive).
Since the the densities $n_{+}\left(q\right)$ and $n_{-}\left(q\right)$
are positive, the RDM diagonal is \emph{never }smaller than its anti-diagonal
and so the ratio $1\ge\Gamma_{1}\left(q,-q\right)/\Gamma_{1}\left(q,q\right)$
is well-defined. The presence of a condensate can perhaps be associated
with a bound of this ratio from below as $q$ grows: 
\begin{equation}
a<\Gamma_{1}\left(q,-q\right)/\Gamma_{1}\left(q,q\right)
\end{equation}

Equality of diagonal and anti-diagonal happens when only even states
are populated! One such case is for the non-interacting Bose gas in
its ground state, where only the (even) ground state is populated,
in this case $D=D_{\text{even}}$ and $D_{\text{odd}}=0$. Once a
non-condensate is formed (due to interactions or increase of temperature,
for example) some of population is transferred into odd states and
therefore $D_{\text{even}}-D_{\text{odd}}$ diminishes. From Eq.~\eqref{eq:anti-diag-integral}
this latter effect causes the reduction of the RDM anti-diagonal integral
$\int\Gamma_{1}\left(q,-q\right)dq$. All the while, the diagonal
integral$\int\Gamma_{1}\left(q,q\right)dq$, remains equal to $D$.
For this reason, a small anti-diagonal population is indicative of
a large non-condensate being formed. 

\bibliographystyle{unsrt}
\bibliography{OmriPaper1}

\end{document}